\input harvmac
\input epsf

\lref\sbook{
R. J. Eden, P. V. Landshoff, D. I. Olive and J. C. Polkinghorne,
{\it The Analytic S-Matrix}, Cambridge University Press, 1966.
}

\lref\ParkeGB{
  S.~J.~Parke and T.~R.~Taylor,
  ``An Amplitude For N Gluon Scattering,''
  Phys.\ Rev.\ Lett.\  {\bf 56}, 2459 (1986).
}

\lref\ManganoXK{
  M.~L.~Mangano, S.~J.~Parke and Z.~Xu,
  ``Duality And Multi-Gluon Scattering,''
  Nucl.\ Phys.\ B {\bf 298}, 653 (1988).
}

\lref\BerendsME{
  F.~A.~Berends and W.~T.~Giele,
  ``Recursive Calculations For Processes With N Gluons,''
  Nucl.\ Phys.\ B {\bf 306}, 759 (1988).
}

\lref\KosowerXY{
  D.~A.~Kosower,
  ``Light Cone Recurrence Relations For QCD Amplitudes,''
  Nucl.\ Phys.\ B {\bf 335}, 23 (1990).
}

\lref\GieleDJ{
  W.~T.~Giele, E.~W.~N.~Glover and D.~A.~Kosower,
  ``Higher order corrections to jet cross-sections in hadron colliders,''
  Nucl.\ Phys.\ B {\bf 403}, 633 (1993)
  [arXiv:hep-ph/9302225].  
}
  
\lref\KunsztMC{
  Z.~Kunszt, A.~Signer and Z.~Trocsanyi,
  ``Singular terms of helicity amplitudes at one loop in QCD and the soft limit
  of the cross-sections of multiparton processes,''
  Nucl.\ Phys.\ B {\bf 420}, 550 (1994)
  [arXiv:hep-ph/9401294].
}

\lref\BernZX{
  Z.~Bern, L.~J.~Dixon, D.~C.~Dunbar and D.~A.~Kosower,
  ``One loop $n$ point gauge theory amplitudes, unitarity and
  collinear limits,''
  Nucl.\ Phys.\ B {\bf 425}, 217 (1994)
  [arXiv:hep-ph/9403226].
}

\lref\BernCG{
  Z.~Bern, L.~J.~Dixon, D.~C.~Dunbar and D.~A.~Kosower,
  ``Fusing gauge theory tree amplitudes into loop amplitudes,''
  Nucl.\ Phys.\ B {\bf 435}, 59 (1995)
  [arXiv:hep-ph/9409265].
}

\lref\BernDB{
  Z.~Bern and A.~G.~Morgan,
  ``Massive Loop Amplitudes from Unitarity,''
  Nucl.\ Phys.\ B {\bf 467}, 479 (1996)
  [arXiv:hep-ph/9511336].
}

\lref\DixonWI{
  L.~J.~Dixon,
  ``Calculating scattering amplitudes efficiently,''
  arXiv:hep-ph/9601359.
}
 
\lref\BernJE{
  Z.~Bern, L.~J.~Dixon and D.~A.~Kosower,
  ``Progress in one-loop QCD computations,''
  Ann.\ Rev.\ Nucl.\ Part.\ Sci.\  {\bf 46}, 109 (1996)
  [arXiv:hep-ph/9602280].
}
  
\lref\BernFJ{
  Z.~Bern, L.~J.~Dixon and D.~A.~Kosower,
  ``Unitarity-based techniques for one-loop calculations in QCD,''
  Nucl.\ Phys.\ Proc.\ Suppl.\  {\bf 51C}, 243 (1996)
  [arXiv:hep-ph/9606378].
} 

\lref\BernSC{
  Z.~Bern, L.~J.~Dixon and D.~A.~Kosower,
  ``One-loop amplitudes for $e^+ e^-$ to four partons,''
  Nucl.\ Phys.\ B {\bf 513}, 3 (1998)
  [arXiv:hep-ph/9708239].
}

\lref\WittenNN{
  E.~Witten,
  ``Perturbative gauge theory as a string theory in twistor space,''
  Commun.\ Math.\ Phys.\  {\bf 252}, 189 (2004)
  [arXiv:hep-th/0312171].
}

\lref\RoibanVT{
  R.~Roiban, M.~Spradlin and A.~Volovich,
  ``A googly amplitude from the B-model in twistor space,''
  JHEP {\bf 0404}, 012 (2004)
  [arXiv:hep-th/0402016].
}

\lref\RoibanKA{
  R.~Roiban and A.~Volovich,
  ``All conjugate-maximal-helicity-violating amplitudes from topological open
  string theory in twistor space,''
  Phys.\ Rev.\ Lett.\  {\bf 93}, 131602 (2004)
  [arXiv:hep-th/0402121].
}

\lref\CachazoKJ{
  F.~Cachazo, P.~Svrcek and E.~Witten,
  ``MHV vertices and tree amplitudes in gauge theory,''
  JHEP {\bf 0409}, 006 (2004)
  [arXiv:hep-th/0403047].
}

\lref\ZhuKR{
  C.~J.~Zhu,
  ``The googly amplitudes in gauge theory,''
  JHEP {\bf 0404}, 032 (2004)
  [arXiv:hep-th/0403115].
}

\lref\RoibanYF{
  R.~Roiban, M.~Spradlin and A.~Volovich,
  ``On the tree-level S-matrix of Yang-Mills theory,''
  Phys.\ Rev.\ D {\bf 70}, 026009 (2004)
  [arXiv:hep-th/0403190].
}

\lref\GukovEI{
  S.~Gukov, L.~Motl and A.~Neitzke,
  ``Equivalence of twistor prescriptions for super Yang-Mills,''
  arXiv:hep-th/0404085.
}

\lref\WuFB{
  J.~B.~Wu and C.~J.~Zhu,
  ``MHV vertices and scattering amplitudes in gauge theory,''
  JHEP {\bf 0407}, 032 (2004)
  [arXiv:hep-th/0406085].
}

\lref\BenaRY{
  I.~Bena, Z.~Bern and D.~A.~Kosower,
  ``Twistor-space recursive formulation of gauge theory amplitudes,''
  Phys.\ Rev.\ D {\bf 71}, 045008 (2005)
  [arXiv:hep-th/0406133].
}

\lref\KosowerYZ{
  D.~A.~Kosower,
  ``Next-to-maximal helicity violating amplitudes in gauge theory,''
  Phys.\ Rev.\ D {\bf 71}, 045007 (2005)
  [arXiv:hep-th/0406175].
}

\lref\BenaXU{
  I.~Bena, Z.~Bern, D.~A.~Kosower and R.~Roiban,
  ``Loops in twistor space,''
  Phys.\ Rev.\ D {\bf 71}, 106010 (2005)
  [arXiv:hep-th/0410054].
}

\lref\CachazoDR{
  F.~Cachazo,
  ``Holomorphic anomaly of unitarity cuts and one-loop gauge theory
  amplitudes,''
  arXiv:hep-th/0410077.
}

\lref\BrittoNJ{
  R.~Britto, F.~Cachazo and B.~Feng,
  ``Computing one-loop amplitudes from the holomorphic anomaly of unitarity
  cuts,''
  Phys.\ Rev.\ D {\bf 71}, 025012 (2005)
  [arXiv:hep-th/0410179].
}

\lref\BernKY{
  Z.~Bern, V.~Del Duca, L.~J.~Dixon and D.~A.~Kosower,
  ``All non-maximally-helicity-violating one-loop seven-gluon amplitudes in
  ${\cal{N}} = 4$ super-Yang-Mills theory,''
  Phys.\ Rev.\ D {\bf 71}, 045006 (2005)
  [arXiv:hep-th/0410224].
}

\lref\BrittoNC{
  R.~Britto, F.~Cachazo and B.~Feng,
  ``Generalized unitarity and one-loop amplitudes in ${\cal N} = 4$
  super-Yang-Mills,''
  Nucl.\ Phys.\ B {\bf 725}, 275 (2005)
  [arXiv:hep-th/0412103].
}

\lref\BernBA{
  Z.~Bern, D.~Forde, D.~A.~Kosower and P.~Mastrolia,
  ``Twistor-inspired construction of electroweak vector boson currents,''
  Phys.\ Rev.\ D {\bf 72}, 025006 (2005)
  [arXiv:hep-ph/0412167].
}

\lref\BernBT{
  Z.~Bern, L.~J.~Dixon and D.~A.~Kosower,
  ``All next-to-maximally helicity-violating one-loop gluon amplitudes in
  ${\cal{N}} = 4$ super-Yang-Mills theory,''
  Phys.\ Rev.\ D {\bf 72}, 045014 (2005)
  [arXiv:hep-th/0412210].
}

\lref\RoibanIX{
  R.~Roiban, M.~Spradlin and A.~Volovich,
  ``Dissolving ${\cal{N}} = 4$ loop amplitudes into QCD tree amplitudes,''
  Phys.\ Rev.\ Lett.\  {\bf 94}, 102002 (2005)
  [arXiv:hep-th/0412265].
}

\lref\BrittoAP{
  R.~Britto, F.~Cachazo and B.~Feng,
  ``New recursion relations for tree amplitudes of gluons,''
  Nucl.\ Phys.\ B {\bf 715}, 499 (2005)
  [arXiv:hep-th/0412308].
}

\lref\BrittoFQ{
  R.~Britto, F.~Cachazo, B.~Feng and E.~Witten,
  ``Direct proof of tree-level recursion relation in Yang-Mills theory,''
  Phys.\ Rev.\ Lett.\  {\bf 94}, 181602 (2005)
  [arXiv:hep-th/0501052].
}

\lref\LuoRX{
  M.~Luo and C.~Wen,
  ``Recursion relations for tree amplitudes in super gauge theories,''
  JHEP {\bf 0503}, 004 (2005)
  [arXiv:hep-th/0501121].
}

\lref\BernHS{
  Z.~Bern, L.~J.~Dixon and D.~A.~Kosower,
  ``On-shell recurrence relations for one-loop QCD amplitudes,''
  Phys.\ Rev.\ D {\bf 71}, 105013 (2005)
  [arXiv:hep-th/0501240].
}

\lref\LuoMY{
  M.~Luo and C.~Wen,
  ``Compact formulas for all tree amplitudes of six partons,''
  Phys.\ Rev.\ D {\bf 71}, 091501 (2005)
  [arXiv:hep-th/0502009].
}

\lref\BedfordYY{
  J.~Bedford, A.~Brandhuber, B.~Spence and G.~Travaglini,
  ``A recursion relation for gravity amplitudes,''
  Nucl.\ Phys.\ B {\bf 721}, 98 (2005)
  [arXiv:hep-th/0502146].
}

\lref\CachazoCA{
  F.~Cachazo and P.~Svrcek,
  ``Tree level recursion relations in general relativity,''
  arXiv:hep-th/0502160.
}

\lref\HodgesBF{
  A.~Hodges,
  ``Twistor diagram recursion for all gauge-theoretic tree amplitudes,''
  arXiv:hep-th/0503060.
}

\newcount\figno
\figno=0 
\def\fig#1#2#3{
\par\begingroup\parindent=0pt\leftskip=1cm\rightskip=1cm\parindent=0pt
\baselineskip=11pt
\global\advance\figno by 1
\midinsert
\epsfxsize=#3
\centerline{\epsfbox{#2}}
\vskip 12pt
{\bf Fig.\ \the\figno: } #1\par
\endinsert\endgroup\par
}
\def\figlabel#1{\xdef#1{\the\figno}}
\def\gb#1{ {\langle #1 ] } }

\Title
{\vbox{
\baselineskip12pt
\hbox{hep-th/0503198}
\hbox{PUPT-2157}
\hbox{NSF-KITP-05-18}
}}
{\vbox{
\centerline{All Split Helicity Tree-Level Gluon Amplitudes}
}}

\centerline{
Ruth Britto${}^\spadesuit$,
Bo Feng${}^\spadesuit$,
Radu Roiban${}^\diamondsuit$,
Marcus Spradlin${}^\clubsuit$,
Anastasia Volovich${}^\clubsuit$}

\bigskip
\bigskip

\centerline{${}^\spadesuit$ School of Natural Sciences, Institute for Advanced
Study, Princeton NJ 08540 USA}

\smallskip

\centerline{${}^\diamondsuit$Department of Physics, Princeton University,
Princeton, NJ 08544 USA}

\smallskip

\centerline{${}^\clubsuit$Kavli Institute for Theoretical Physics,
University of California,
Santa Barbara, CA 93106 USA}

\bigskip
\bigskip

\centerline{\bf Abstract}

Recently a new recursion relation for tree-level gluon amplitudes
in gauge theory has been discovered.  We solve this recursion
to obtain explicit formulas for
the closed set of amplitudes with arbitrarily many positive
and negative helicity gluons in a split helicity configuration.
The solution admits a 
simple diagrammatic expansion in terms
of `zigzag' diagrams. We comment on generalizations of this result.

\bigskip

\Date{March 2005}

\newsec{Introduction}

Gluon scattering amplitudes are important for computing
jet processes as backgrounds in hadron colliders.
However, the number of Feynman diagrams required in calculating these
amplitudes quickly exceeds practical bounds as the number of external
gluons increases.
Remarkably, the final form of these amplitudes is often far simpler than
one would guess from the expansion in Feynman diagrams.  At tree level,
the first examples of such simplicity came in the work of Parke and
Taylor \ParkeGB, who found an elegant single-term expression for
maximally helicity violating (MHV) amplitudes.
This formula was proven in \ManganoXK\ using the Berends-Giele recursion
\BerendsME.

In a remarkable paper \WittenNN\ Witten found that
gluon amplitudes are localized on certain curves in twistor space
and proposed a twistor string theory capturing the properties of these
amplitudes.
This discovery enabled a deeper understanding of their structure
\refs{\RoibanVT,\RoibanKA, \CachazoKJ,\ZhuKR,\RoibanYF,\KosowerYZ,\GukovEI,
\WuFB,\BenaRY,\BernBA}.  One particular approach \refs{\CachazoKJ} offered a
new diagrammatic expansion of tree amplitudes in terms of MHV vertices.
This was a great improvement over the number and computational complexity
of Feynman diagrams.

Another development in understanding the structure of gauge theory
amplitudes has come from reconsidering their analytic structure in
the space of complexified momenta \sbook\ and making use of the
interplay between tree and loop amplitudes.
The divergent behavior of the one-loop contribution to a scattering
amplitude encodes the tree-level contribution \refs{\GieleDJ,\KunsztMC}.
Following recent progress in computing ${\cal N}=4$ amplitudes at
one-loop \refs{\BenaXU,\CachazoDR,\BrittoNJ,\BrittoNC,\BernBT}
using unitarity based methods \refs{\BernZX,\BernCG,
\BernSC,\BernFJ,\BernDB,\BernJE}
this singular behavior was used in \refs{\BernKY,\BernBT,\RoibanIX} to
derive new representations of tree amplitudes.  In particular, a study of
analytic structure revealed that one-loop
${\cal N}=4$ amplitudes could be expressed essentially as a sum of products
of tree amplitudes times known functions \BrittoNC.  From this result,
combined with a
new formula derived from
the  singular behavior relations, it was inferred \RoibanIX\ that one
could express a tree amplitude of gluons in terms of pairs of tree
amplitudes of gluons and adjoint fermions and scalars with fewer external
legs. In \BrittoAP, it was proposed
that this inference could be modified to involve only gluon amplitudes.
An explicit quadratic recursion relation was conjectured and was found
to yield directly the most compact formulas for tree amplitudes
known so far.

The recursion relation of \BrittoAP\ was first proven and properly
understood in \BrittoFQ\ by considering the analytic properties of
amplitudes in the space of complexified momenta.  It was also shown
that the recursion could be generalized and applied to prove the
validity of the MHV diagrams of \CachazoKJ.
More recently it has been found that the recursion relation can be
translated into terms of  twistor geometry \HodgesBF, and
there have been extensions to
amplitudes with fermions \refs{\LuoRX,\LuoMY} or gravitons
\refs{\BedfordYY,\CachazoCA} and
loop amplitudes \BernHS.

In this paper we study tree-level gluons amplitudes in helicity
configurations of the form $({-}{-}\cdots{-}{+}{+}\cdots{+})$ which
we call `split helicity amplitudes.'
These amplitudes form a closed
set under the recursion relation (as well as under collinear limits).
Here we solve the recursion to derive a simple expression for
any split helicity amplitude.
For an amplitude with $q$ negative helicity gluons and $n-q$ positive
helicity gluons, our expression has ${n - 4 \choose q - 2}$ terms.
Each term has
can be interpreted as a `zigzag diagram'
involving the cyclic arrangement of gluons on a closed curve.

It would be very interesting to
understand whether amplitudes other than those with
split helicity  configurations
can be captured by some generalization of
these zigzag diagrams.
It would also be interesting to investigate the relationship of these
diagrams to
the diagrams of twistor geometry, for which some related results have
appeared in \HodgesBF.

\newsec{Preliminaries}

It is convenient to write tree amplitudes of gluons in the
spinor-helicity formalism.
In four dimensions any null vector $p$
can be written as a bispinor, $p_{a\dot a} =
\lambda_a\widetilde\lambda_a$, where
$\lambda_a$ and $\widetilde\lambda_{\dot a}$ are spinors of positive and
negative chirality respectively.
The inner product of vectors can be
written in terms of the natural inner product of spinors
$\vev{i~j} = \epsilon_{ab}\lambda_i^a\lambda_j^b$ and
$[i~j] = \epsilon_{\dot a\dot
b}\widetilde\lambda^{\dot a}_i\widetilde\lambda^{\dot b}_j$.\foot{Our convention
for the sign of $[i~j]$ follows \WittenNN\ and is
the opposite of the convention in much of the earlier physics literature.
In particular, $(p_i + p_j)^2 = \vev{i~j} [i~j] = - \langle i|p_j|i]$ here.}
We work with amplitudes of cyclically ordered gluons, and all momenta are
directed outward.  For further details and references, see
\refs{\WittenNN,\DixonWI}.

Sums of cyclically consecutive momenta will be denoted by
\eqn\aaa{
P_{x,y} \equiv p_x + p_{x+1} + \cdots + p_y~.
}
We define the products
\eqn\ppou{\eqalign{
\gb{i|P_{x,y} |j} & \equiv \sum_{k=x}^y \vev{i~k}[k~j]~,\cr
\vev{i|P_{x_1,y_1}\cdots P_{x_r,y_r} |j} & \equiv
\sum_{k_1=x_1}^{y_1} \cdots \sum_{k_r=x_r}^{y_r}
\vev{i~k_1}[k_1~k_2]\cdots\vev{k_r~j}~.
}}

The recursion relation of \BrittoAP\ can be written as follows.
\eqn\retrue{\eqalign{ & A(1^-,2,\ldots , (n-1),n^+) = \cr &
\qquad\qquad\qquad\sum_{i=2}^{n-2}\sum_{h=+,-} \left ( A({\widehat 1},2,\ldots,
i,-{\widehat P}^h_{1,i} ) {1\over P^2_{1,i}} A(+{\widehat
P}^{-h}_{1,i}, i+1,\ldots , n-1, {\widehat n} ) \right), } }
where
\eqn\deff{ \eqalign{
\widehat P_{1,i} & = P_{1,i} +{P_{1,i}^2\over \gb{1|P_{1,i}|n}}
\lambda_{1} \tilde\lambda_{n}~, \cr \widehat p_{1} & =  p_{1}
+{P_{1,i}^2\over \gb{1|P_{1,i}|n}} \lambda_{1}
\tilde\lambda_{n}~, \cr \widehat p_{n} & =  p_{n} -{P_{1,i}^2\over
\gb{1|P_{n,i}|n}} \lambda_{1} \tilde\lambda_{n}~. } }
Here we have chosen the gluons labeled by $1$ and $n$ to be the
reference gluons.

It was noted in \BrittoAP\ that amplitudes in split helicity
configurations are closed under the recursion and that for this case,
the sum \retrue\ has only two nonzero terms.
We turn our attention to such amplitudes in the next section.
However, some of the technical
points appearing in our calculation below turn out to be generic to the
application of \retrue\ to any class of amplitudes, and we will comment
on such generalizations when appropriate.


\newsec{Split Helicity Amplitudes and Zigzag Diagrams}

The main result of this note is the
following formula for general tree-level split helicity gluon amplitudes:
\eqn\fullzig{
A(1^-,\ldots,q^-,(q{+}1)^+,\ldots,n^+) =
\sum_{k=0}^{\min(q-3,n-q-2)} \sum_{A_k,B_{k+1}} {N_1 N_2 N_3 \over D_1 D_2
D_3}~.
}
Here $A_k$ and $B_{k+1}$ respectively range over all subsets of
the indices $\{2,\ldots,q-2\}$ 
and $\{q+1,\ldots,n-1\}$
of cardinality $k$ and $k+1$.
In increasing numerical order, the elements are labeled
$a_1,a_2,\ldots,a_k$ and $b_{k+1},\ldots,b_1$.
There are a total of ${n - 4 \choose q - 2}$ terms in the sum.
The quantities $N$ and $D$ are defined by
\eqn\pieces{\eqalign{
N_1 &= {\langle 1 | P_{2,b_1} P_{b_1 + 1, a_1} P_{a_1 + 1, b_2} \cdots
P_{b_{k+1} + 1,q - 1}|q\rangle^3}~,\cr
N_2 &=
\langle b_1{+}1~b_1 \rangle \langle b_2{+}1~b_2 \rangle
\cdots
\langle b_{k+1}{+}1~b_{k+1} \rangle~,\cr
N_3 &= [a_1~a_1{+}1] \cdots [a_k~a_k{+}1]~,\cr
D_1 &=
P^2_{2,b_1} P^2_{b_1 + 1, a_1} P^2_{a_1 + 1, b_2} \cdots
P^2_{b_{k+1} + 1,q - 1}~,\cr
D_2 &= F_{q,1} \overline{F}{}_{2,q-1}~,\cr
D_3 &= [2|P_{2,b_1}|b_1{+}1 \rangle \langle b_1|P_{b_1 + 1,a_1}|a_1]
[a_1{+}1|P_{a_1+1,b_2}|b_2{+}1\rangle
\cdots
\langle b_{k+1}|P_{b_{k+1}+1,q-1}|q{-}1]~,
}}
where  $F_{x,y}$ is given by
\eqn\aaa{
F_{x,y} = \langle x~x{+}1\rangle \langle x{+}1~x{+}2\rangle
\cdots \langle y{-}1~y\rangle~,
}
and
$\overline{F}_{x,y}$ is
given by the same expression but with the inner product
$[\cdot~\cdot]$.

We find it helpful to illustrate each term of \fullzig\ by a zigzag diagram,
drawn as follows:
$$
\figins{\vcenter{\epsfxsize230pt\epsfbox{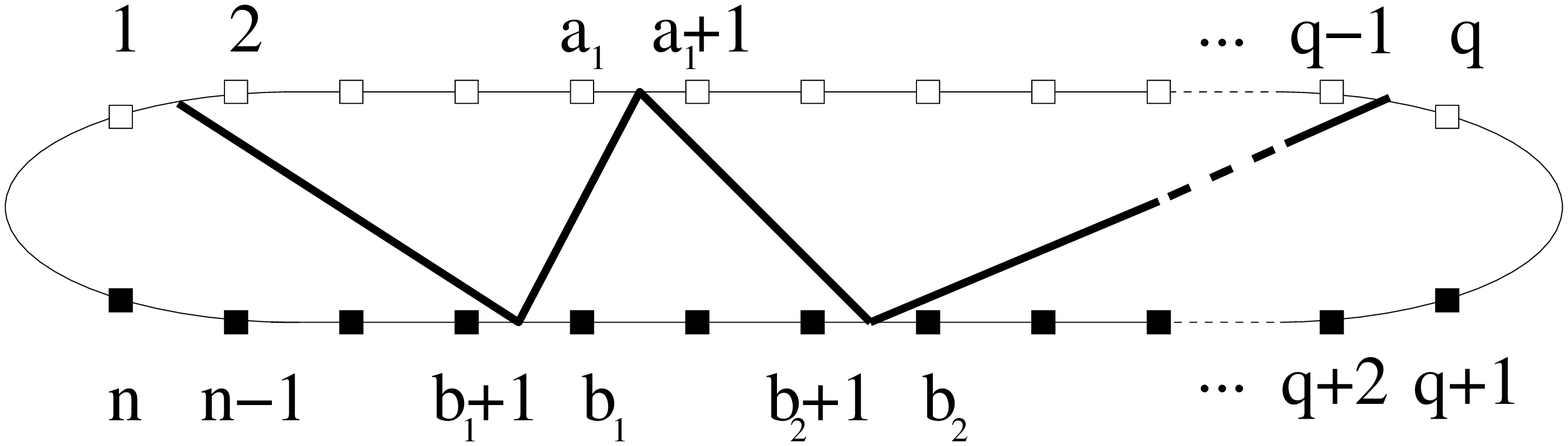}}}~.
$$
Arrange the gluon indices in clockwise order around a closed curve,
with the negative helicities $\{1,\ldots,q\}$ on the top side and the
positive helicities $\{q+1,\ldots,n\}$ on the bottom side.
A zigzag is a connected collection of non-self-intersecting line 
segments which begins at $(1,2)$ and ends at $(q-1,q)$, alternating
at each step between the top and bottom sides.
It is clear that there is a one-to-one correspondence between such
zigzag diagrams and choices of the subsets $A_k$ and $B_{k+1}$.
The line segments in a zigzag diagram are in one-to-one correspondence with the
momenta $P_{x,y}$ appearing in the expressions \pieces,
and the rule
for transforming any given zigzag diagram into a formula is clear from \pieces.

The observation
that the line segments correspond
to cyclic sums of momenta motivates a relatively simple diagrammatic proof
that \fullzig\ has the correct multi-particle factorization property.
In particular,
contributions to the $1/P_{x,y}^2$ pole in $A_{q,n-q}$ can only come
from those zigzags which contain a line segment that cuts the amplitude
so that gluons $(x,x+1,\ldots,y)$ are on one side of the cut.


\newsec{Examples}

In this section we illustrate the
application of  \fullzig\ by means of several
examples.

\subsec{MHV amplitudes}

We can view the MHV amplitudes ($q=2$) as a special case where
the zigzag collapses to a point.  In this case there are no $P$'s,
so the only factors in \pieces\ which contribute are $N_1 = \langle
1~2\rangle^3$ and $D_2 = F_{2,1}$.  This gives immediately
the desired result
\eqn\mhv{
A(1^-,2^-,3^+,\ldots,n^+)=
\figins{\vcenter{\epsfxsize80pt\epsfbox{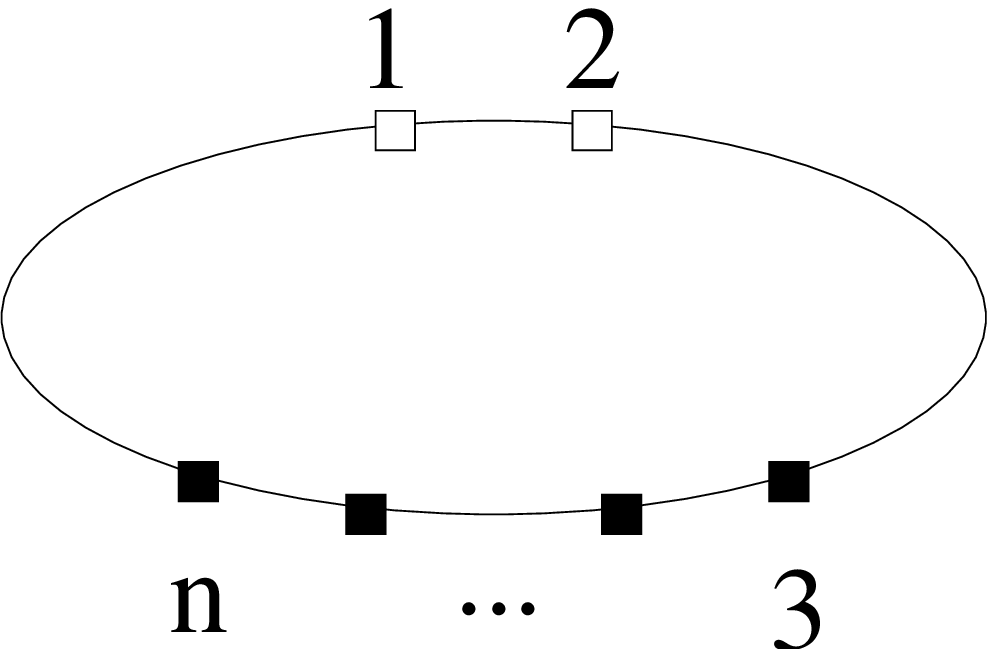}}}
= { \langle 1~2\rangle^3 \over \langle 2~3\rangle \langle 3~4 \rangle
\cdots \langle n~1\rangle}~.
}

\subsec{$\overline{\rm MHV}$ amplitudes}

For ${\overline{\rm MHV}}$ amplitudes ($n=q+2$) there is only
a single zigzag diagram,
\eqn\nmhv{\eqalign{
&A(1^-,2^-,\ldots,q^-,(q{+}1)^+,(q{+}2)^+)=
\figins{\vcenter{\epsfxsize80pt\epsfbox{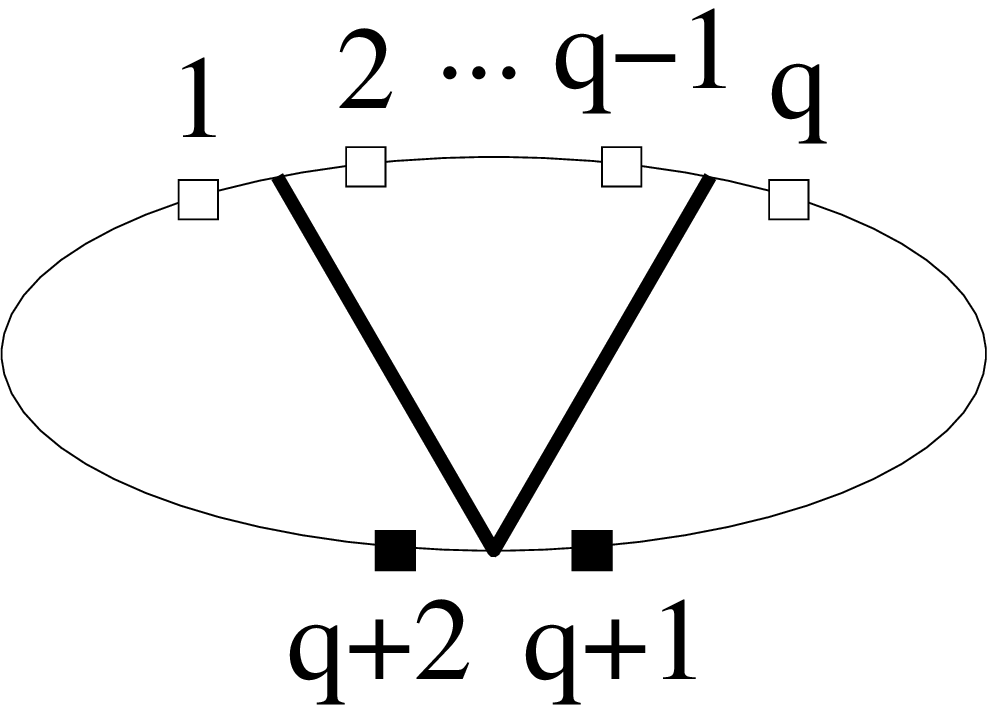}}}\cr
&=
{1 \over F_{q,1} \overline{F}{}_{2,q-1}}
{ \langle 1 | P_{2,q+1} P_{q+2,q-1}|q\rangle^3 
\over P_{2,q+1}^2 P_{q+2,q-1}^2
}
{ \langle q{+}2~q{+}1 \rangle \over
[2|P_{2,q+1} |q{+}2\rangle\langle q{+}1 | P_{q+2,q-1}|q-1]}~,
}}
which immediately simplifies to the expected result
\eqn\aaa{
A(1^-,2^-,\ldots,q^-,(q{+}1)^+,(q{+}2)^+)=
{ [q{+}1~q{+}2]^3 \over [q{+}2~1][1~2] \cdots [q~q{+}1]}~.
}

The zigzag rules treat MHV
and ${\overline{\rm MHV}}$ amplitudes differently.
More generally, the zigzag rules do not manifestly respect the
symmetry
\eqn\aaa{
A_{q,n-q}(1^-,\ldots,q^-,(q{+}1)^+,\ldots,n^+)
= \overline{A_{n-q,q}((q{+}1)^-,\ldots,n^-,1^+,\ldots,q^+)}~.
}
It is clear that one can formulate an alternate set of
rules involving zigzags that begin and end on the positive helicity
side of the diagram, with some straightforward changes to \pieces.
This alternate set of rules might
lead to simpler intermediate expressions for amplitudes which have
more negative than positive helicity gluons, although the final
expression for any amplitude would of course
be equal to that obtained from the version of the rules that we presented
above.

\subsec{Split helicity next-to-MHV amplitudes}

Next we consider the split helicity next-to-MHV amplitudes ($q=3$) for
arbitrary $n$.  In this case there is only one kind of zigzag diagram
which contributes, allowing us to immediately write down the result
\eqn\aaa{\eqalign{
A(1^-,2^-,3^-,4^+,\ldots,n^+)&=\sum_{j=4}^{n-1}
\figins{\vcenter{\epsfxsize80pt\epsfbox{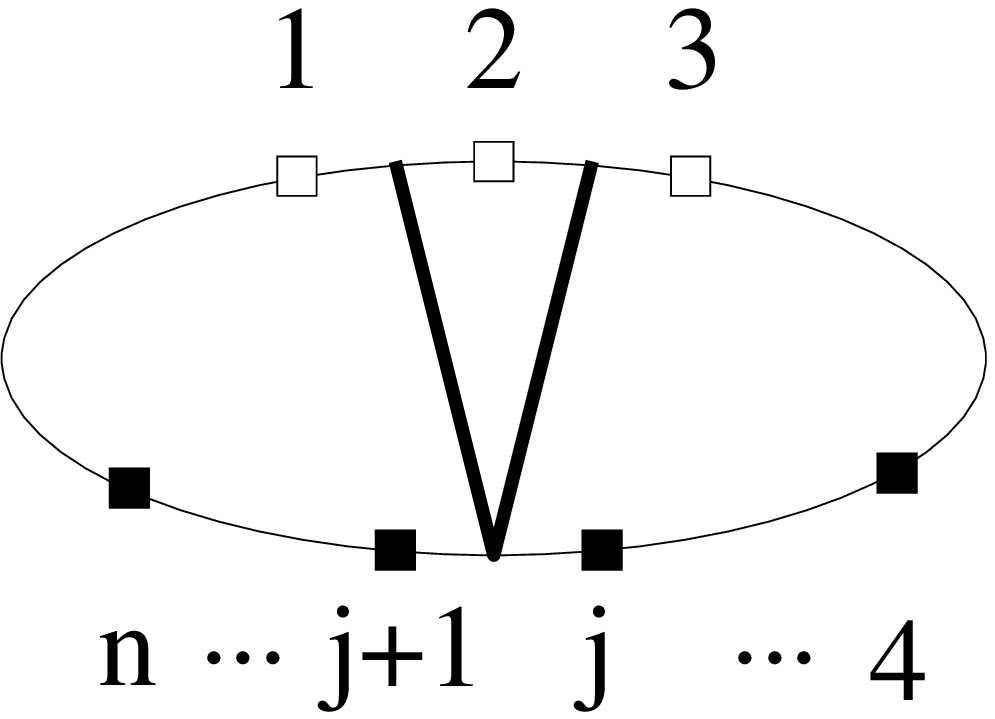}}}\cr
&=
{1 \over F_{3,1}}
\sum_{j=4}^{n-1} {
\langle 1 | P_{2,j} P_{j+1,2}|3\rangle^3 \over
P_{2,j}^2 P_{j+1,2}^2}
{\langle j{+}1~j \rangle \over
[2|P_{2,j}|j{+}1\rangle \langle j|P_{j+1,2}|2]}~.
}}
This formula, which is equivalent to one recently obtained in \LuoRX,
is noticeably more compact than previously-known expressions
for next-to-MHV amplitudes \refs{\KosowerXY,\CachazoKJ,\KosowerYZ}.

\subsec{The amplitude $A(1^-,2^-,3^-,4^-,5^+,6^+,7^+,8^+)$}

A compact six-term formula for 
the eight-particle split helicity next-to-next-to-MHV amplitude
$A(1^-,2^-,3^-,4^-,5^+,6^+,7^+,8^+)$
was first written down in \refs{\RoibanIX}.
It is straightforward to check that this formula is immediately reproduced
by summing the following six zigzag diagrams:

$$
\figins{\vcenter{\epsfxsize75pt\epsfbox{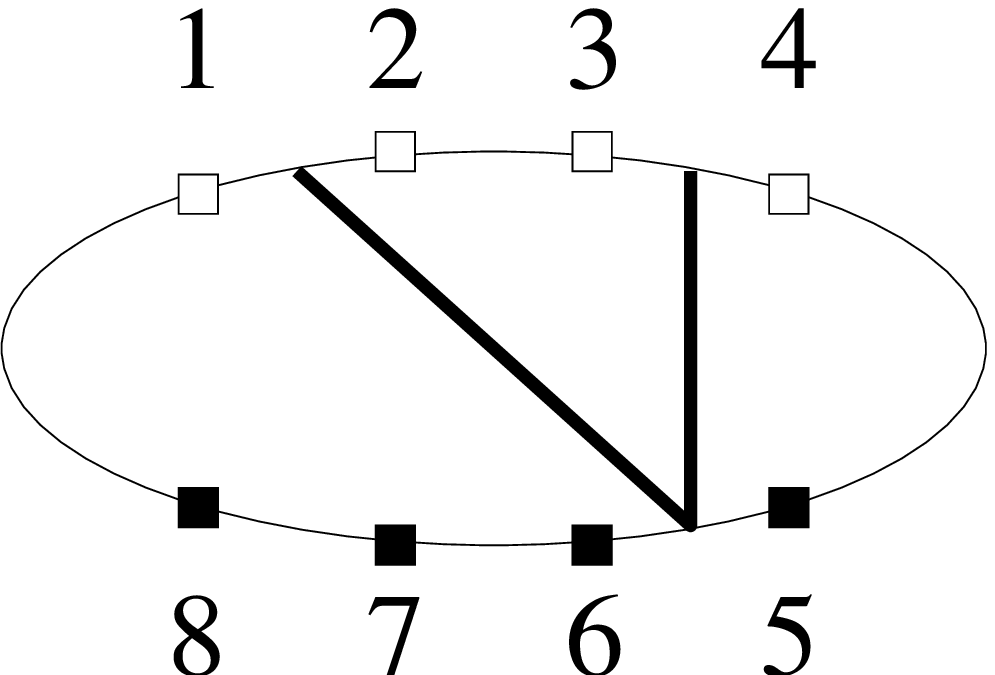}}} =
{1 \over F_{4,1} \overline{F}{}_{2,3}}
{\langle 1 | P_{2,5} P_{6,3}|4\rangle^3 \over P_{2,5}^2 P_{6,3}^2 }
{ \langle 6~5 \rangle \over [2|P_{2,5}|6\rangle \langle 5|P_{6,3}|3]}~,
$$

$$
\figins{\vcenter{\epsfxsize75pt\epsfbox{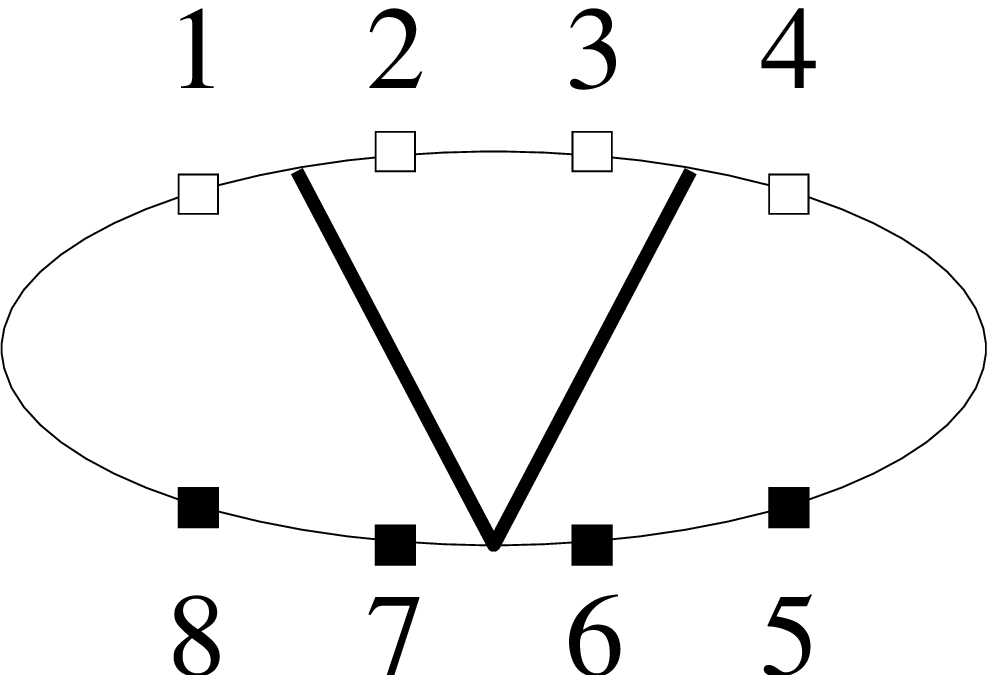}}} =
{1 \over F_{4,1} \overline{F}{}_{2,3}}
{\langle 1 | P_{2,6} P_{7,3}|4\rangle^3 \over P_{2,6}^2 P_{7,3}^2 }
{ \langle 7~6 \rangle \over [2|P_{2,6}|7\rangle \langle 6|P_{7,3}|3]}~,
$$

$$
\figins{\vcenter{\epsfxsize75pt\epsfbox{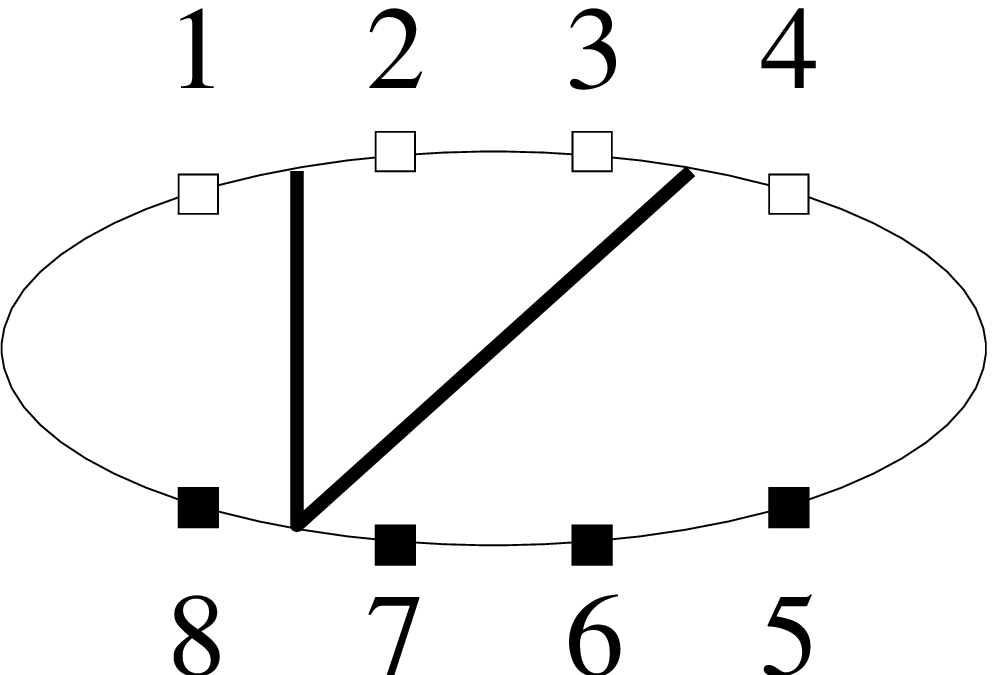}}} =
{1 \over F_{4,1} \overline{F}{}_{2,3}}
{\langle 1 | P_{2,7} P_{8,3}|4\rangle^3 \over P_{2,7}^2 P_{8,3}^2 }
{ \langle 8~7 \rangle \over [2|P_{2,7}|8\rangle \langle 7|P_{8,3}|3]}~,
$$

$$
\figins{\vcenter{\epsfxsize75pt\epsfbox{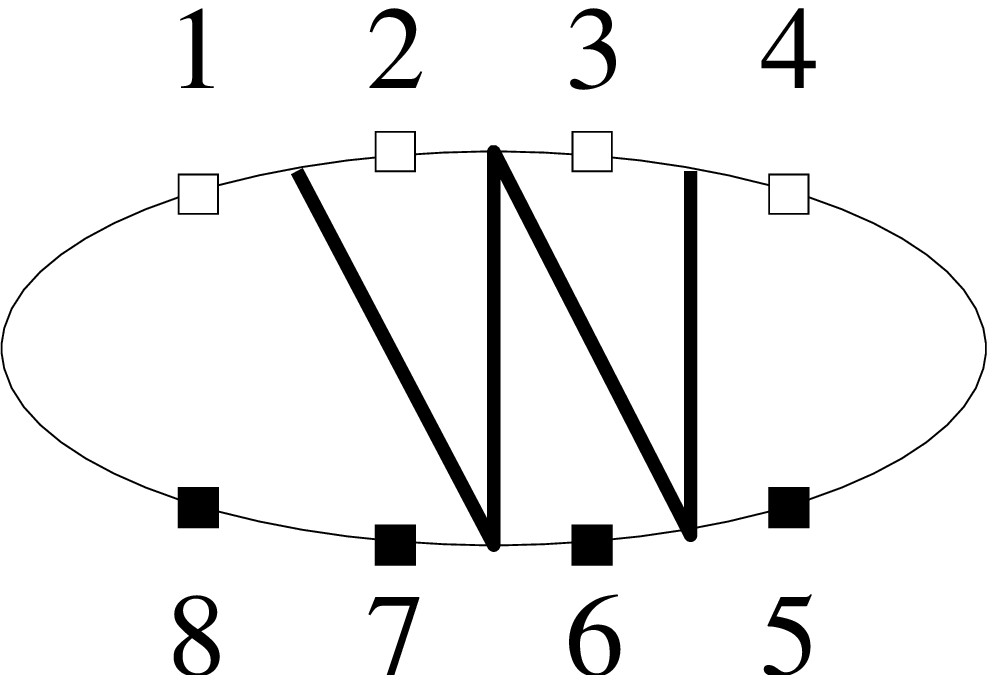}}} =
{1 \over F_{4,1} \overline{F}{}_{2,3}}
{ \langle 1 | P_{2,6} P_{7,2} P_{3,5} P_{6,3} |4\rangle^3 \over
P^2_{2,6} P^2_{7,2} P^2_{3,5} P^2_{6,3}}
{\langle 7~6 \rangle
\qquad
[2~3]
\qquad
\langle 6~5 \rangle
\over
[2|P_{2,6}|7\rangle \langle 6|P_{7,2}|2]
[3|P_{3,5}|6\rangle \langle 5|P_{6,3}|3]}~,
$$

$$
\figins{\vcenter{\epsfxsize75pt\epsfbox{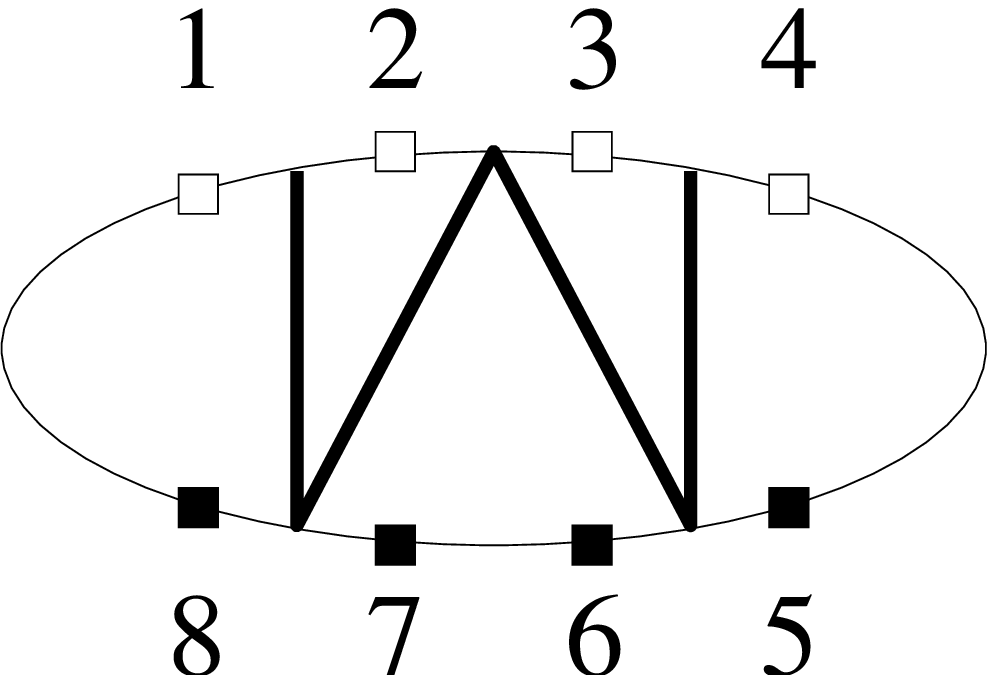}}} =
{1 \over F_{4,1} \overline{F}{}_{2,3}}
{ \langle 1 | P_{2,7} P_{8,2} P_{3,5} P_{6,3} |4\rangle^3 \over
P^2_{2,7} P^2_{8,2} P^2_{3,5} P^2_{6,3}}
{\langle 8~7 \rangle
\qquad
[2~3]
\qquad
\langle 6~5 \rangle
\over
[2|P_{2,7}|8\rangle \langle 7|P_{8,2}|2]
[3|P_{3,5}|6\rangle \langle 5|P_{6,3}|3]}~,
$$

$$
\figins{\vcenter{\epsfxsize75pt\epsfbox{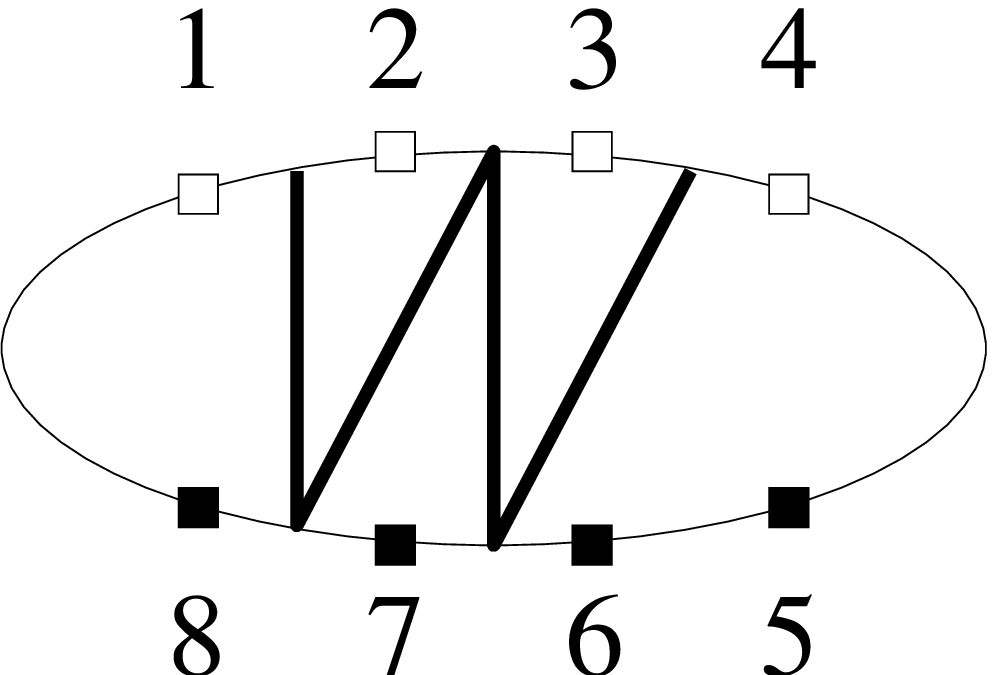}}} =
{1 \over F_{4,1} \overline{F}{}_{2,3}}
{ \langle 1 | P_{2,7} P_{8,2} P_{3,6} P_{7,3} |4\rangle^3 \over
P^2_{2,7} P^2_{8,2} P^2_{3,6} P^2_{7,3}}
{\langle 8~7 \rangle
\qquad
[2~3]
\qquad
\langle 7~6 \rangle
\over
[2|P_{2,7}|8\rangle \langle 7|P_{8,2}|2]
[3|P_{3,6}|7\rangle \langle 6|P_{7,3}|3]}~.
$$

We have written out each term carefully in order to emphasize
how simple it is to translate each picture into the corresponding
formula, but of course several of the terms can be somewhat simplified.

\newsec{Proof of the Main Result}

In this section we present an elementary proof of \fullzig.
We first reduce the quadratic
recursion \retrue\ to a simpler linear version.
This processed version of the recursion relates an amplitude with $q$
negative helicity gluons to amplitudes with only $q-1$ negative helicity
gluons.  This can be done for any kind of amplitude, but for split
helicity amplitudes the resulting recursion turns out to be simple enough
that it can easily be solved by induction in the
number of negative
helicity gluons.
The initial condition for the induction is the case $q=2$,
which in section 4.1 we showed gives the correct result.
In what follows we use the notation
\eqn\aaa{
A_{q,n-q} = A_{q,n-q}(1^-,\ldots,q^-,(q{+}1)^+,\ldots,n^+)~.
}

\bigskip

\noindent
{\bf Step 1.}

First we apply the recursion \retrue\ to the amplitude
$A_{q,n-q}$ with reference momenta $n$ and $1$.
There are only two nonvanishing contributions.
In each term, one of the amplitudes is a three-gluon vertex
and the other is an $n-1$ gluon amplitude.
Specifically, we find (after a little
simplification) that \retrue\ can be cast into the form
\eqn\eqone{
A_{q,n-q} =
{[n~2] \over [n~1] [1~2]}
A_{q-1,n-q}(\widehat{2}^-,\ldots,\widehat{n}^+)
+ {\langle n{-}1~1 \rangle \over \langle n{-}1~ n \rangle
\langle n~1 \rangle}
A_{q,(n-1)-q}(\widehat{1}^-,\ldots,\widehat{n{-}1}{}^+)~,
}
where the shifted spinors are
\eqn\aaa{\eqalign{
\lambda_{\widehat{1}} &= \lambda_1~,\cr
\lambda_{\widehat{2}} &= \lambda_2 + {[n~1] \over [n~2]}\lambda_1~,\cr
\lambda_{\widehat{n{-}1}} &= \lambda_{n{-}1}~,\cr
\lambda_{\widehat{n}} &= \lambda_n + {[1~2] \over [n~2]} \lambda_1~,\cr
\widetilde{\lambda}{}_{\widehat{1}} &= \widetilde{\lambda}{}_1
+ { \langle n{-}1~n \rangle \over \langle n{-}1~1\rangle}
\widetilde{\lambda}{}_n~,\cr
\widetilde{\lambda}{}_{\widehat{2}} &= \widetilde{\lambda}{}_2~,\cr
\widetilde{\lambda}{}_{\widehat{n{-}1}} &= \widetilde{\lambda}{}_{n{-}1}
+ {\langle n~1 \rangle \over \langle n{-}1~1\rangle} \widetilde{\lambda}{}_n~,\cr
\widetilde{\lambda}{}_{\widehat{n}} &= \widetilde{\lambda}{}_n~.
}}
Note that we have relabeled the momentum $\widehat{P}$ to $\widehat{2}$
in the first term in \eqone\ and to $\widehat{n{-}1}$ in the second term.
This puts \eqone\ into a form that can easily be fed back into itself.

\bigskip

\noindent
{\bf Step 2.}

Next we would like to use \eqone\ to express $A_{q,n-q}$ in terms
of amplitudes which have strictly less than $q$ negative helicity gluons.
We leave the first term in \eqone\ alone since it already has $q-1$
negative helicity gluons, but to the second term we apply the
result of \eqone\ again to rewrite (schematically)
\eqn\aaa{
A_{q,(n-1)-q} = A_{q-1,(n-1)-1} + A_{q,(n-2)-q}~.
}
We continue applying \eqone\ in this manner to strip away $j$ positive
helicity gluons from the amplitude.
It is very convenient to continue using the first and last gluons
(in the order written) as the reference gluons.
This process terminates
at $j=n-q-1$ since an amplitude with only a single positive
helicity gluon vanishes, and we obtain the desired expression
\eqn\eqtwo{
A_{q,n-q} = - \sum_{j=0}^{n-q-2}
{\langle 1|P_{2,n-j-1}|2] \over F_{n-j,1} P_{2,n-j-1}^2 \langle
n{-}j|P_{2,n-j-1}|2]}
A_{q-1,(n-j)-q}(\widehat{2}^-,\ldots,\widehat{n{-}j}{}^+)~,
}
in terms of the shifted spinors
\eqn\aaa{\eqalign{
\lambda_{\widehat{2}} &=  { P_{n-j,2}
P_{2,n-j-1} |1 \rangle
\over \langle 1|P_{2,n-j-1}|2]}~,\cr
\lambda_{\widehat{n{-}j}} &= { P_{2,n-j-1}|2] \over
\langle 1|P_{2,n-j-1} |2]}~,\cr
\widetilde{\lambda}{}_{\widehat{2}} &= \widetilde{\lambda}{}_2~,\cr
\widetilde{\lambda}{}_{\widehat{n{-}j}} &=
P_{2,n-j-1}|1\rangle~.
}}

It is important to note that
the procedure of expressing an amplitude with  
$q$ negative helicity gluons in terms of
amplitudes with only $q-1$ negative
helicity gluons
works in complete generality.  Let us use
$A_{n_1^-, n_1^+, \ldots,n_k^-,n_k^+}$ to denote
an amplitude in which the first
$n_1^-$ gluons have negative helicity, the next $n_1^+$ have
positive helicity, and so on.  By repeatedly 
splitting off the last group of positive helicity gluons, in a manner 
similar to what we did above, it is  possible
to show that the
recursion \retrue\ can always be processed into the form
\eqn\aaa{
A_{n_1^-, n_1^+, \ldots,n_k^-,n_k^+} = \sum_{j=0}^{n_k^+ - 2}
\left[
U_j(p) A_{(n_1-1)^-, n_1^+, \ldots,n_k^-,(n_k-j)^+} + R_j(p)\right]~.
}
The functions $U_j(p)$ are universal functions of
the spinors, depending
only on the three-gluon amplitudes $({+}{+}{-})$ and
$({+}{-}{-})$.
In \eqtwo\ we have determined these functions from the analysis
of the special case $k=1$.
In contrast, 
the $R_j(p)$ term depends sensitively on the particular class of amplitudes
under consideration.
It is the happy fact that $R_j(p)$ vanishes for split helicity amplitudes
that will allow us to solve the recursion \eqtwo\ inductively in this
case.

\bigskip

\noindent
{\bf Step 3.}

This step is nothing but convenient bookkeeping:
we simply relabel the summation index in \eqtwo\ from
$j$ to $b = n - j - 1$. After some other minor changes,
we have
\eqn\eqthree{
A_{q,n-q} = \sum_{b = q+1}^{n-1}
{\langle 1|P_{2,b}|2] \over F_{b+1,1} P_{b+1,1}^2 \langle
b{+}1|P_{b+1,1} |2]}
A_{q-1,(b+1)-q}(\widehat{2}^-,\ldots,\widehat{b{+}1}{}^+)~,
}
with shifted spinors
\eqn\shiftthree{\eqalign{
\lambda_{\widehat{2}} &=  {
P_{3,b} P_{b+1,1}|1\rangle
\over \langle 1|P_{2,b}|2]}~,\cr
\lambda_{\widehat{b{+}1}} &= { P_{2,b}|2] \over
\langle 1|P_{2,b} |2]}~,\cr
\widetilde{\lambda}{}_{\widehat{2}} &= \widetilde{\lambda}{}_2~,\cr
\widetilde{\lambda}{}_{\widehat{b{+}1}} &=
P_{2,b}|1\rangle~.
}}

\bigskip

\noindent
{\bf Step 4.}

The aim of our proof is now to 
show that the expression \fullzig\ inductively follows
from the processed recursion
\eqthree.
Let us start by explaining how the zigzag diagrams which contribute
to $A_{q-1,(b+1) - q}$ on the right-hand side of \eqthree\ lift
to zigzag diagrams contributing to $A_{q,n-q}$.
For fixed $b \in \{q+1,\ldots,n-1\}$, these diagrams fall into two
categories.  Either $b_1$ (the largest element of the subset $B_k$)
is equal to $b$, or it is less than $b$.
In the former case, the lift looks like

\eqn\typeone{
\figins{\vcenter{\epsfxsize11.2cm\epsfbox{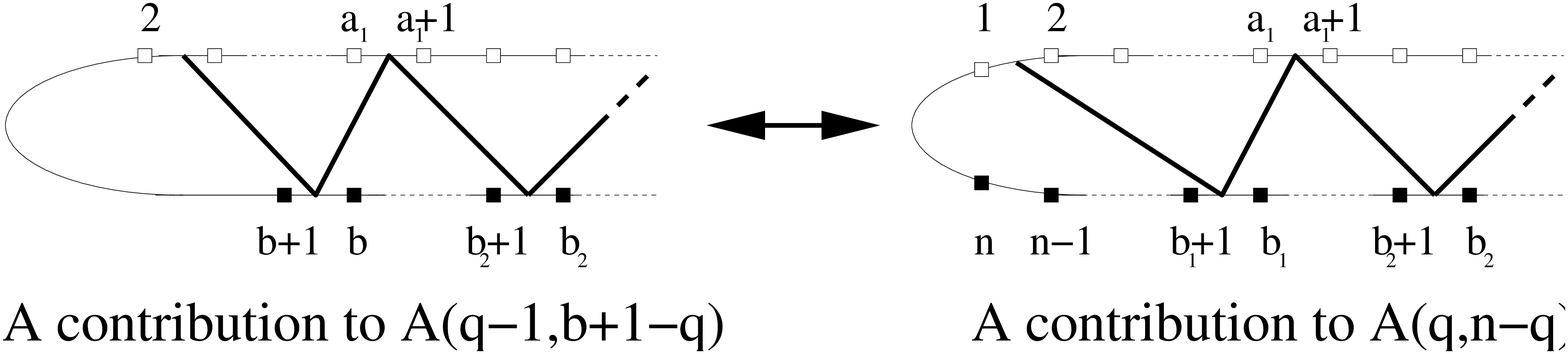}}}
}

\noindent
while in the latter case, the lift involves the addition of an extra zigzag
(shown here as a dotted line for emphasis)

\eqn\typetwo{
\figins{\vcenter{\epsfxsize11.2cm\epsfbox{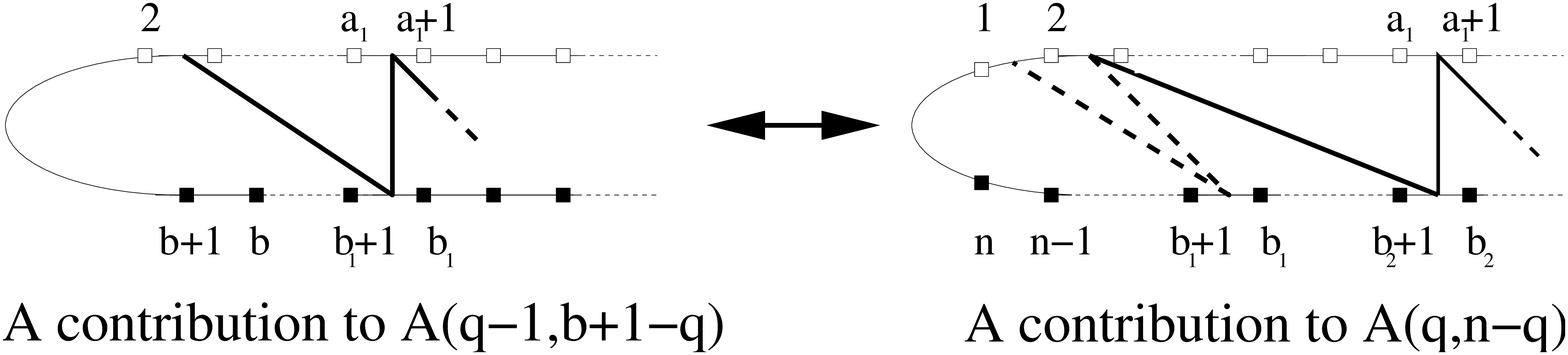}}}
}

\bigskip

The reverse map is also clear:  a zigzag diagram contributing
to $A_{q,n-q}$ comes from the $b = b_{1}$ or the $b = b_2$ term in the
sum \eqthree\ depending on whether $a_1 > 2$ or $a_1 = 2$ respectively.
We have therefore established a one-to-one correspondence between the 
zigzag diagrams appearing on the left-hand side of \eqthree\ and those
appearing on the right-hand side.  
The final step is to establish the quantitative agreement between
the two sides.

\bigskip

\noindent
{\bf Step 5a.}
First consider, for fixed $b \in \{q+1,\ldots,n+1 \}$, a zigzag
diagram contributing to $A_{q-1,(b+1)-q}$ which has $b_{1} = b$.
Note that none of the $P_{x,y}$ appearing carry a single hatted
index ($\widehat{2}$ or $\widehat{b{+}1}$).  They always appear together
as
\eqn\aaa{
\cdots + p_{b} + p_{\widehat{b{+}1}} + p_{\widehat{2}} + p_3 + \cdots
= \cdots + p_b + p_{b{+}1} + p_{b{+}2} + \cdots + p_2 + p_3 + \cdots~,
}
so we never need to worry about any hats appearing on the $P$'s.
Using \shiftthree\ we find
\eqn\aaa{
\eqalign{
N_1 &= \langle \widehat{2} | P_{3,b} P_{b+1,a_1} \cdots P_{b_{k+1}+1,q-1}|q
\rangle^3
= \left[
{ (P_{3,b}^2)^3 \over \langle 1 | P_{2,b}|2]^3}
\right]
\langle 1|P_{2,b} P_{b+1,a_1} \cdots P_{b_{k+1}+1,q-1}|q
\rangle^3~,\cr
N_2 &= \langle \widehat{b{+}1}~b\rangle
\langle b_2~b_2{+}1\rangle
\cdots
\langle b_{k+1}{+}1~b_{k+1}\rangle
\cr
&= 
\left[
{1 \over F_{b,b+1}}
{\langle b|P_{2,b}|2]
\over \langle 1|P_{2,b}|2]}\right]
\langle b{+}1~b\rangle
\langle b_2~b_2{+}1\rangle
\cdots
\langle b_{k+1}{+}1~b_{k+1}\rangle~,\cr
N_3 &= [a_1~a_1{+}1] \cdots [a_k~a_k{+}1]~,\cr
D_1 &= P_{q,b_1}^2 P_{b_1+1,a_1}^2 P_{a_1+1,b_2}^2 \cdots
P_{a_k+1,b}^2
P_{b+1,2}^2\cr
&= \left[{P_{b+1,2}^2 \over P_{b+1,1}^2} \right]
P_{q,b_1}^2 P_{b_1+1,a_1}^2 P_{a_1+1,b_2}^2 \cdots P_{a_k+1,b}^2
P_{b+1,1}^2~,\cr
D_2 &= F_{q,\widehat{2}} \overline{F}{}_{3,q-1}
=\left[ {\langle
b|P_{2,b}|2]\over \langle 1 |P_{2,b}|2]^2}
P_{3,b}^2 {1 \over F_{b,1} [2~3]}
\right]
F_{q,1} \overline{F}{}_{2,q-1}~,\cr
D_3 &=
[3|P_{3,b}|\widehat{b{+}1}\rangle \langle b|P_{b+1,a_1} |a_1]
\cdots \langle b_{k+1}|P_{b_{k+1}+1,q-1}|q{-}1]\cr
&=\left[ {P_{3,b}^2
[2~3]
\over \langle 1 |P_{2,b} |2] \langle b{+}1|P_{b+1,2}|2]}
\right]
[2|P_{3,b}|b{+}1\rangle \langle b|P_{b+1,a_1} |a_1] \cdots
\langle b_{k+1}|P_{b_{k+1}+1,q-1}|q{-}1]~.
}} 
In each term we have assembled in brackets
all of the extra factors which
are not present in the corresponding contribution to the zigzag diagram
on the right-hand side of \typeone.
When we take the ratio $N_1 N_2 N_3/D_1 D_2 D_3$ we find that these
extra factors combine into
\eqn\aaa{
{ F_{b+1,1} P_{b+1,1}^2 \langle b{+}1|P_{b+1,1}|2]
\over \langle 1|P_{2,b}|2]}~,
}
which precisely cancels the factor appearing in \eqthree.
This establishes the
quantitative agreement between the diagrams shown in \typeone.

\smallskip

\noindent
{\bf Step 5b.}
Finally, we consider the case shown in \typetwo, with a zigzag
diagram contributing to $A_{q-1,(b+1)-q}$ such that $b_{1} < b$.
In this case the analysis is simpler, since only $N_1$ 
and $D_2$ develop extra factors relative to what one would expect
from looking at the second diagram in \typetwo:
\eqn\aaa{\eqalign{
N_1 &=
\langle \widehat{2}|P_{3,b_1} P_{b_1 + 1,a_1} \cdots P_{b_{k+1}+1,q-1}
|q\rangle^3\cr
&= \left[ {1 \over \langle 1| P_{2,b} |2]^3}
\right]
\langle 1|P_{2,b_1} P_{b_1 + 1,a_1} \cdots P_{b_{k+1}+1,q-1}
|q\rangle^3~,\cr
D_2 &= F_{q,\widehat{2}} \overline{F}{}_{3,q-1}
=\left[-{\langle
b|P_{2,b}|2]\over \langle 1 |P_{2,b}|2]^2}
P_{3,b}^2 {1 \over F_{b,1} [2~3]}
\right]
F_{q,1} \overline{F}{}_{2,q-1}~.
}}
Note that $N_1$ has grown the two additional 
legs of the zigzag, shown as dotted lines in \typetwo.
Combining the factors in brackets with the factors
appearing explicitly in \eqthree\ gives
\eqn\aaa{
{\langle b{+}1~b\rangle
[2~3] \over P_{3,b}^2
P_{b+1,1}^2 
[2|P_{2,b}|b\rangle
\langle b{+}1| P_{b+1,1}|2]}~,
}
which are precisely the correct extra factors needed to account for
the addition of the dotted zigzag.

\vfill

\centerline{\bf Acknowledgments}\nobreak

We would like to thank Z.~Bern and F.~Cachazo for helpful discussions and L.~Dixon for pointing out typos in equations in sections 3 and 4.4 of the previous version.
This research was supported in part by the Department of Energy under Grant
No. DE-FG02-91ER40671 (RR) and by the
National Science Foundation under
Grants Nos.~PHY99-07949 (MS, AV) and PHY-0070928 (RB, BF).

\listrefs

\end